\documentclass[11pt,twoside]{article}

\usepackage{asp2006}
\usepackage{epsf}
\usepackage{psfig}
\usepackage{lscape}

\begin{document}
\title{Arp102B: An ADAF and a Torus ?}   
\author{Ranga-Ram Chary}   
\affil{Spitzer Science Center, MS220-6, Caltech, Pasadena, CA 91125}    

\begin{abstract} 

Arp102B is a nearby radio galaxy which displays the presence
of double peaked Balmer emission lines. Sub-arcsec Keck mid-infrared
imaging and Spitzer spectroscopy reveal a spatially compact mid-infrared
source which displays tentative evidence for variability. The
F$_{\nu}\propto\nu^{-1.2}$ spectral energy distribution is suggestive of an advection
dominated accretion flow. The absence of dust features over the 5$-$40$\mu$m range
make it unlikely that thermal dust emission dominates the mid-infrared luminosity.
We also detect the presence of molecular hydrogen in emission which is 
asymmetrically redshifted by $\sim$500-1000 km/s from the 
systemic velocity of the galaxy. 
Since the forbidden, low ionization lines in this galaxy are 
at the systemic velocity,
we suggest that the molecular hydrogen emission arises from a
rotating molecular gas structure surrounding the nuclear black hole at 
a distance of $\sim$1~pc.

\end{abstract}


\section{Introduction}   
Arp 102B is an E0, radio loud galaxy at a distance of 104.9 Mpc \citep{Eracleous:04}. 
Stellar
dynamical measurements indicate a velocity dispersion of 188$\pm$8 km~s$^{-1}$ \citep{Barth:02}.
which implies a black hole mass of $\sim$10$^{8}$~M$_{\sun}$ \citep[See also][]{Newman:97}. 
The AGN in
Arp 102B is of the ``double peaked" emitter class. 
The simplest possible explanation for this spectral profile identifies an 
accretion disk as the source of these lines. The AGN also shows strong low ionization lines and very
weak/absent high ionization lines \citep{Stauffer:83, Halpern:96}. 
The weakness of the high
ionization lines and absence of double peaked structure in them supports the accretion disk origin
under the assumption that the outer parts of the thin disk are invisible to the photoionizing source.
The absence of a strong UV bump, presence of a hard X-ray source and 
low Eddington luminosity
of this object are indicative of an advection dominated flow \citep[ADAF;][]{Ho:00}.

In this paper, we present the results from high spatial resolution mid-infrared imaging
and moderate resolution infrared spectroscopy of the nucleus of Arp102B.
Mid-infrared fine structure emission lines (e.g. [NeII], [NeIII], [OIV])
are less affected by extinction and help determine 
if shocks or photoionization are responsible for
the line excitation mechanism. The mid-infrared is also the regime of molecular H$_{2}$ rotational
lines which provide an independent measure of temperature and mass of the surrounding gas. 

\section{Broad-band properties}   
In the Keck/LWS observations undertaken in 2000, the AGN in Arp102B was found to have a flux density of 90$\pm$20 mJy at 12.5$\mu$m.
The source was undetected at 17.9$\mu$m with a 3$\sigma$ flux density limit of 50 mJy.
In the Spitzer observations in 2005, the source has decreased in brightness by more than a factor of 2 at
12$\mu$m since the Keck
observations and has shown
a change in the infrared spectral energy distribution (Figure 1). 
The Spitzer measurements are also consistent with the \citet{Puschell:86}
value of 23$\pm$7 mJy derived from 10.6$\mu$m observations made in 1981. As a result, it is unclear if the Keck data
suffer from photometric calibration errors or if there actually is a factor of $\sim$2 variability in the brightness of the source.

The broadband flux density in the Spitzer data can be best represented by 
the form F$_{\nu}$(mJy)=1.7$\times\lambda^{1.23\pm0.15}$
with evidence for a 
turnover at $\lambda>20\mu$m. Such a spectrum is remarkably similar to that observed in NGC 4258 \citep{Chary:00} where 
the emission was thought to arise from self-Comptonized synchrotron radiation. 
The 5$-$40 $\mu$m luminosity of the source is 6.7$\times$10$^{9}$~L$_{\sun}$ which is comparable to the 0.2-10 keV
luminosity measured by ROSAT and ASCA in 1991 and 1998 respectively \citep{Eracleous:03}.
The bolometric luminosity derived by integrating over the full SED 
including the radio/millimeter measurements by \citet{Puschell:86} is 8$\times$10$^{43}$~ergs~s$^{-1}$. This implies that the Eddington 
ratio of the 10$^{8}$~M$_{\sun}$
black hole is $\sim$6$\times$10$^{-3}$, about 5 times larger than previous estimates and marginally within
the ADAF limit.

\section{Emission Line properties}
\subsection{Low Ionization Lines}
Figure 2 shows the emission lines that are detected in the spectrum of Arp 102B.
As noted from optical/UV spectroscopy, the spectrum of Arp102B is dominated by low ionization metallic
fine structure lines with a notable absence of high ionization lines. 
[OIV] which has an ionization
potential of 55 eV is the most prominent of these features but is known to exist in starburst galaxies. 
Adopting the simplified model of \citet{Sturm:02}, the [OIV]/[NeII] ratio of 0.21$\pm$0.05 
and absence of high ionization lines would indicate that that 90\% of the bolometric luminosity in
Arp102B is powered by a starburst. The measured [FeII]/[OIV] ratio of 0.86$\pm$0.25 is also remarkably
similar to that of NGC 6240 which is thought to be a starburst dominated source \citep{Lutz:03}.

However, Arp 102B shows an absence of strong polycyclic aromatic hydrocarbon dust emission
or silicate features. PAH are commonly thought to be
present in starbursts. It is unclear if the PAH are 
absent because they are destroyed in the hard radiation field around an AGN/super starburst or because
there is no dust in the vicinity of the AGN. The latter is a less likely option given the presence 
of molecular gas emission in the infrared spectrum.

With a [NeII]/[SIII] ratio of 2.3$\pm$0.5
and a [NeIII]/[ArII]$>3.1$, Arp102B is out of the starburst powered regime and intermediate between the
shock excited and photoionization excited regimes. The pure photoionization models in
\citet{Spin:92} reproduce the 
observed [NeIII]/[OIII] $\lambda$5007 ratios very well but underestimate the [NeII]/[NeIII] ratio. 
Ionizing shocks from starbursts and supernova remnants, can account for the observed [FeII]/[OIV]
ratio as well as the high [NeII] line flux. Thus, we conclude
that the low ionization line spectrum of Arp102B
is a composite, dominated by photoionization but with $\sim$20-30\%
contribution from slow shocks.

\subsection{Molecular Hydrogen Emission}
One of the striking features of the spectrum are the S1 and S3 rotational lines of H$_{2}$ which appear
to be offset in velocity from the systemic velocity of the galaxy. The ratio of the S(3)/S(1) line fluxes
is 1.6 indicative of warm molecular gas at a temperature of $\sim$400 K. 
None of the other H$_{2}$ lines are detected. 
Although the forbidden
lines are observed at the systemic velocity of the galaxy, the H$_{2}$ lines are offset by $>2\sigma$
at both the orders. This is unlikely to be due to wavelength calibration uncertainties. We suggest
that the molecular hydrogen might be in a rotating molecular gas structure with the redshifted emission
stronger than the blue shifted emission due to a warp. We do not detect
the blue shifted emission of the S(3) line because of inadequate wavelength coverage in the spectrum.
The size of the structure is about $\sim$1 pc, ranging from 0.5$-$2.7 pc depending on
the adopted velocity range.

Although the mid-infrared lines are offset, the near-infrared H$_{2}$ lines in ground-based spectra
appear to be at the systemic velocity of the galaxy \citep{Rodri05}. This is surprising. One possibility is that the
near-infrared continuum and line emission is being dominated by the starlight while the mid-infrared
is being dominated by the region around the AGN. Furthermore, the strength of the near-infrared
lines is far too strong for the gas temperature that we derive from the ratio of the mid-IR H$_{2}$ lines.
Thus, unless there are unknown wavelength calibration uncertainties in the mid-IR spectrum,
there must be a hotter gas component which dominates the near-IR molecular gas emission.


\acknowledgements 

I would like to acknowledge my collaborators, Luis Ho, Harlan Devore,
Eric Becklin and Nick Scoville for their assistance with different
aspects of the observations. I would also like to acknowledge the organizers for a 
very enjoyable and productive meeting. These observations were partly a component  
of the Spitzer Teacher's Education program. 


\begin{figure}
\plotfiddle{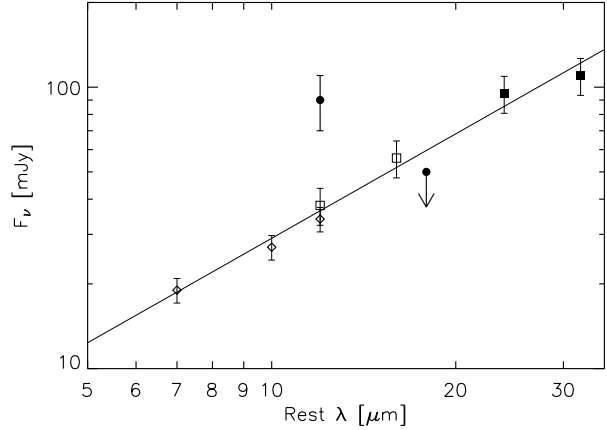}{2in}{0}{50}{50}{-170}{-190}
\caption{Infrared spectral energy distribution of Arp102B from Spitzer 2005 data and Keck 2000 data (solid circles).
The luminosity of the nuclear source has decreased by a factor of 2.5 at 12$\mu$m between the Keck and Spitzer observations
while its spectral shape has inverted, suggestive of a transition from a thin disk to an ADAF. However, calibration
uncertainties are a possible source of error for the Keck data.}
\end{figure}

\begin{figure}[b]
\plottwo{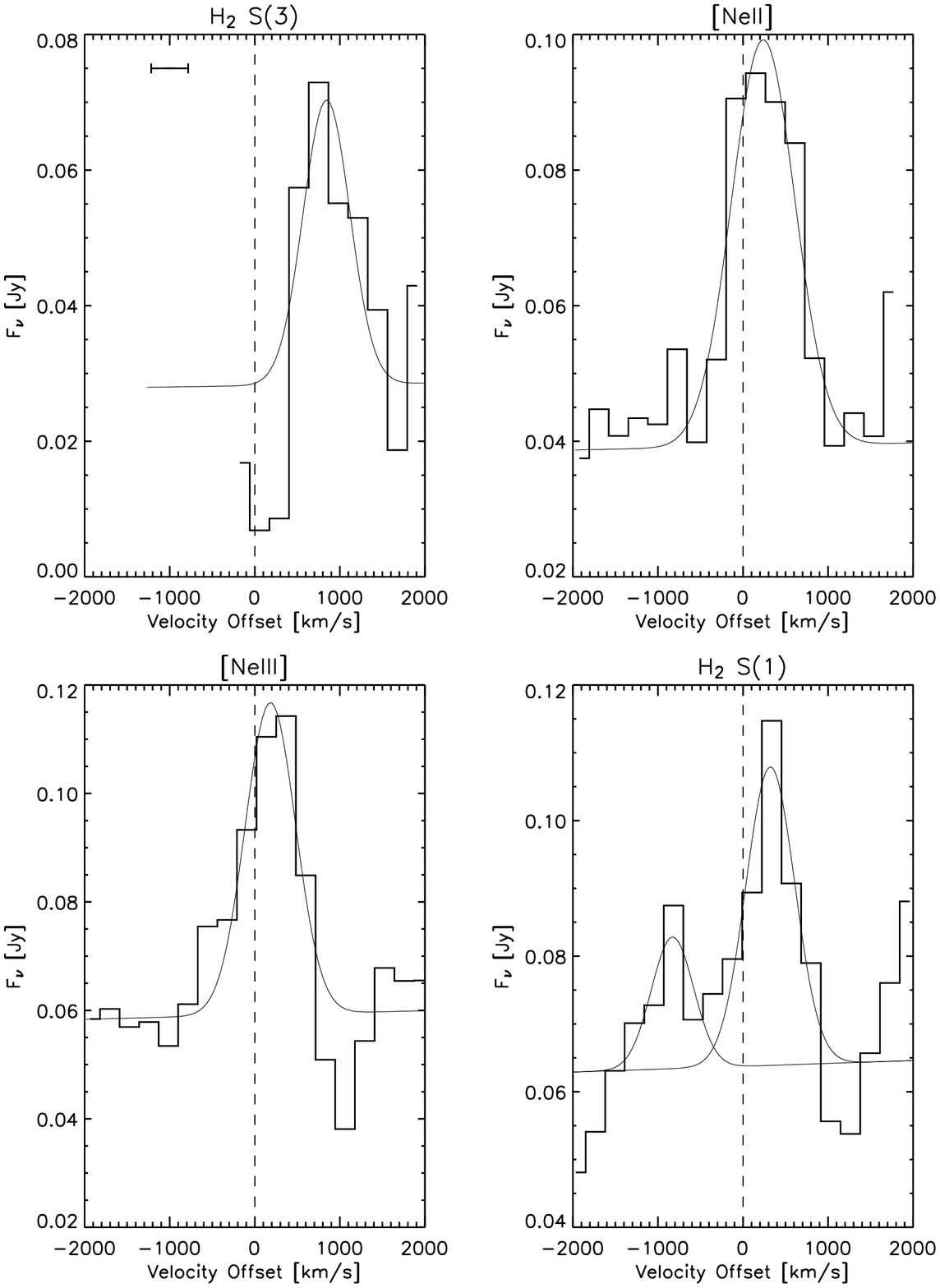}{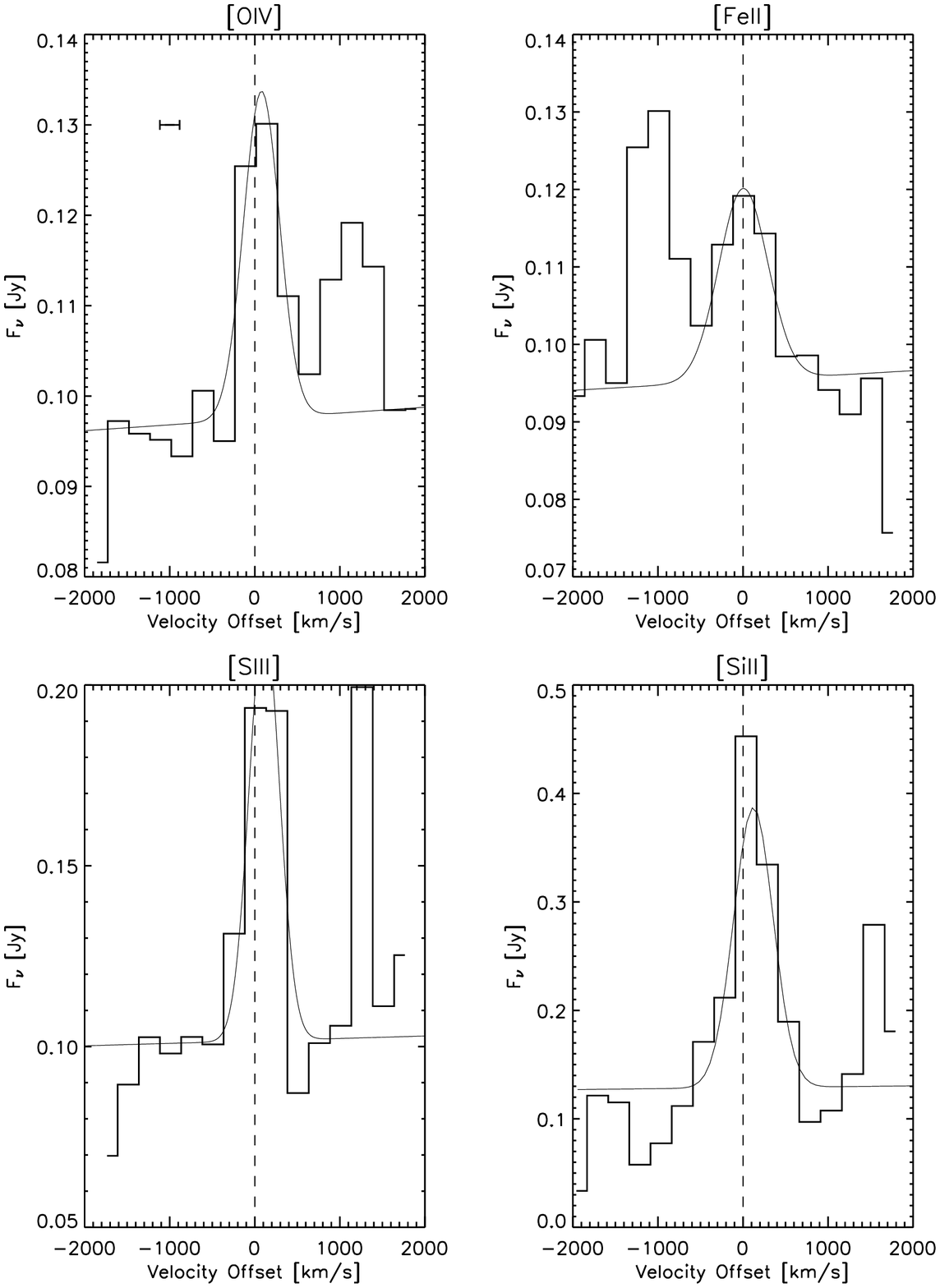}
\caption{Spectral lines seen in the Spitzer/IRS spectra with a best fit Gaussian and constant baseline. Also
shown is the $\pm$1$\sigma$ uncertainty in the wavelength calibration for the SH and LH modules. Although the 
forbidden lines are consistent with no velocity offset to within the calibration uncertainty, the H$_{2}$ lines
are offset by 500$-$1000 km~s$^{-1}$ from the systemic velocity. Given the mass of the black hole, it implies
the molecular hydrogen is $\sim$1 pc from the nucleus.}
\end{figure}

\end{document}